# Design and Implementation of Flight Visual Simulation System

Feng Tian[1], Wenjian Chai[1], Chuanyun Wang[1],

[1] School of Computer Science, Shenyang Aerospace University,
110136 Shenyang, China
{tianfeng5861, cimu.love, wangcy0301}@163.com

**Abstract.** The design requirement for flight visual simulation system is studied, and the overall structure and development process are proposed in this paper. Through the construction of 3D scene model library and aircraft model, the rendering and interaction of visual scene are implemented. The changes of aircraft flight attitude in visual system are controlled by real-time calculation of aircraft aerodynamic and dynamic equations and flight simulation effect is enhanced by this kind of control. Several key techniques for optimizing 3D model and relative methods for large terrain modeling are explored for improving loading ability and rendering speed of the system. Experiment shows that, with specific function and performance guaranteed as a premise, the system achieves expected results, that is, precise real-time calculation of flight attitude and smooth realistic screen effect.

## 1  Introduction

The Flight Visual Simulation System (FVSS) is an important integral part of flight simulator. The system can provide vivid 3D scenes and effective flight information including realistic flight environment and flight attitude. Flight equipment operations can be quickly, safely and skillfully mastered by using the system. Building an independent FVSS can reduce the cost and development circle of flight simulator. Internal and unforeseen faults within aircraft body or caused by complex flight environment can be avoided through the FVSS [1].

Based on the previous studies of visual simulation, an overall design and implementation of flight visual simulation system is proposed in this paper. Techniques of model optimization and large terrain modeling in 3D modeling are described and the simulation accuracy and authenticity of flight attitude are enhanced. For PC environment, generally speaking, the virtual reality technique based visual simulation system has better performance in aspects such as visual immersion and authenticity. However, simulation of flight motion is not adequate. Although advanced computer languages such as C++ and so on are used for simulation implementation of basic flight kinetic equations, the implemented system has complex code, long development cycle, low reliability and difficult maintenance [2]. In order to overcome these shortages, simulation software, Matlab, is used for calculation of flight kinetic equations; generated results are exported to Visual Studio 7 for mix programming in order to improve authenticity of simulated flight motion. The proposed system has realistic 3D models and high immersion virtual scenes.

## 2    Design and Development Process of FVSS

The FVSS contains two main modules, flight simulation and visual simulation. For flight simulation module, the main focus is on the modeling of flight motion equations including aircraft aerodynamic and dynamic equations. The 3D motion status of aircraft in flight simulation module is controlled and updated continuously in accordance with the calculation results of these equations. And for visual simulation module, distinct flight scene is constructed through simulations of terrain, physiognomy, ground buildings and weather conditions. Main structure of the system is shown in Fig.1.

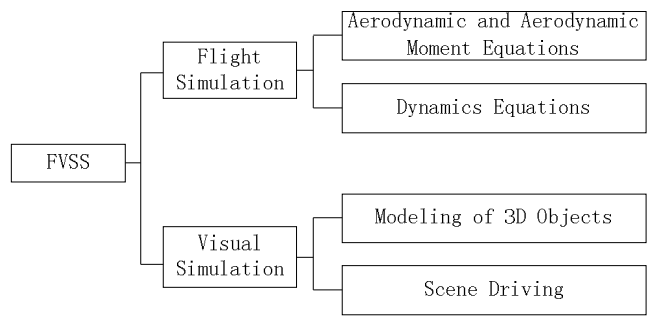

**Fig. 1.** Main structure of the FVSS

In order to accelerate development speed, assure system quality and improve system maintenance, software Creator and Vega Prime from MultiGen-Paradigm are used for the development of visual simulation module. Calculation for flight simulation module is carried out by Matlab and generated results are exported to Visual Studio 7 for mix programming which implements the real-time control of aircraft model flight motion in Vega prime. Actual development process is described by the following steps and the process is depicted in Fig.2.

Step1: Software Creator is used for 3D modeling and flight motion simulation is created by light kinetic equations;

Step2: Configuration, run and rendering of flight scene are through the call for .acf file in Vega Prime;

Step3: Parameters of aircraft in visual simulation, such as flight attitude, velocity, acceleration and so on, are calculated by Matlab.

Step4: .exe file is generated by using the integrated simulation platform of Visual Studio 2003. In this platform, generated data from Matlab are used for real-time connection between flight attitude and visual scene of aircraft model in Vega Prime. The visual simulation module and flight simulation module are combined together through the real-time connection and control.

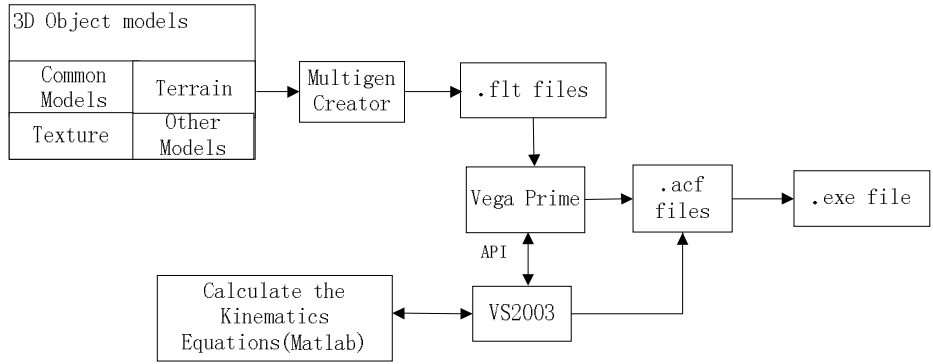

**Fig. 2.** Actual development process of the FVSS

## 3  Mathematical Modeling of Flight Simulation

The aircraft flight motion simulation, as an important part of FVSS, directly affects the reliability and authenticity of the system. Flight motion simulation effect can be greatly improved by relative mathematical models of aircraft flight dynamics. In this paper, the FVSS is based on two assumptions:

   a. Flight area is the space above ground level where the rotation of earth and the curvy motion of mass center of earth are neglected.
   b. Aircraft is an ideal rigid body and influence from aircraft body elastic deformation and rotating parts are not considered [3].

### 3.1  Models of Aerodynamic and Aerodynamic Moment

When an aircraft is in the air, besides gravity and motor power, aerodynamic forces, drag X, lift Y and lateral compression Z which are all caused by relative airflow also act on the aircraft. Rotation inertia exists moment of inertia and would generate three moments; rolling moment $M_x$, yawing moment $M_y$ and pitching moment $M_z$. These aerodynamic forces and moments are as follows [4]:

   Where, $C_x$, $C_y$ and $C_z$ are coefficient of drag, lift and lateral compression separately; $m_x$, $m_y$ and $m_z$ are coefficients of rolling moment, yawing moment and pitching moment separately; q is speed pressure, S is wing area, b is wingspan and c is average aerodynamic chord.

$$\begin{aligned} X &= qSC_x \\ Y &= qSC_y \\ Z &= qSC_z \end{aligned} \quad (1)$$

$$\begin{aligned} M_x &= qSbm_x \\ M_y &= qSbm_y \\ M_z &= qScm_z \end{aligned} \quad (2)$$

### 3.2 Models of Dynamics

During the process of simulation, the 6-DOF motion equation is used for flight motion calculation. The calculation generated parameters for both body coordinate system and inertial coordinate system are axis acceleration, velocity, position, angular acceleration, angular velocity, attitude angle, attack angle, side slip angle, flight path pitch angle and flight path azimuth angle.

Following is the dynamics equation group of mass center motion in aircraft body axis system [5].

$$\begin{bmatrix} m\left(\dfrac{dV_x}{dt} + \omega_y V_z - \omega_z V_y\right) \\ m\left(\dfrac{dV_y}{dt} + \omega_z V_x - \omega_x V_z\right) \\ m\left(\dfrac{dV_z}{dt} + \omega_x V_y - \omega_y V_x\right) \end{bmatrix} = \begin{bmatrix} F_x + g_x \\ F_y + g_y \\ F_z + g_z \end{bmatrix} \quad (3)$$

Where, $V_x$, $V_y$ and $V_z$ are components of flight acceleration in aircraft body axis system; $\omega_x$, $\omega_y$ and $\omega_z$ are components of aircraft angular velocity; $F_x$, $F_y$ and $F_z$ are the components of total force in aircraft body axis system.

$$\begin{aligned} J_x \dfrac{d\omega_x}{dt} - (J_y - J_z)\omega_y\omega_z - J_{yz}(\omega_y^2 - \omega_z^2) - J_{zx}\left(\dfrac{d\omega_z}{dt} + \omega_x\omega_y\right) - J_{xy}\left(\dfrac{d\omega_y}{dt} - \omega_z\omega_x\right) &= M_x \\ J_y \dfrac{d\omega_y}{dt} - (J_z - J_x)\omega_z\omega_x - J_{xz}(\omega_z^2 - \omega_x^2) - J_{xy}\left(\dfrac{d\omega_x}{dt} + \omega_y\omega_z\right) - J_{yz}\left(\dfrac{d\omega_z}{dt} - \omega_x\omega_y\right) &= M_y \\ J_z \dfrac{d\omega_z}{dt} - (J_x - J_y)\omega_x\omega_y - J_{xy}(\omega_x^2 - \omega_y^2) - J_{yz}\left(\dfrac{d\omega_y}{dt} + \omega_z\omega_x\right) - J_{xz}\left(\dfrac{d\omega_x}{dt} - \omega_y\omega_x\right) &= M_z \end{aligned} \quad (4)$$

And above is the dynamics equation group for aircraft rotation about center of mass. Where, $J_x$, $J_y$ and $J_z$ are inertia moment; $J_{xy}$, $J_{yz}$ and $J_{xz}$ are product of inertia; $\omega_x$, $\omega_y$ and $\omega_z$ are angular velocities about each aircraft body axis separately and $M_x$, $M_y$ and $M_z$ are resultant moments about each axis separately.

Due to the symmetry of aircraft, $J_{yz}$ and $J_{xz}$ can be treated as 0, so above equation group can be simplified to the following form:

$$J_x \frac{d\omega_x}{dt} - (J_y - J_z)\omega_y\omega_z - J_{xy}\left(\frac{d\omega_y}{dt} - \omega_z\omega_x\right) = M_x$$
$$J_y \frac{d\omega_y}{dt} - (J_z - J_x)\omega_z\omega_x - J_{xy}\left(\frac{d\omega_x}{dt} + \omega_y\omega_z\right) = M_y \quad (5)$$
$$J_z \frac{d\omega_z}{dt} - (J_x - J_y)\omega_x\omega_y - J_{xy}\left(\omega_x^2 - \omega_y^2\right) = M_z$$

The relationship between change rate of attitude angle and component of angular velocity can be expressed by the following equations:

$$\dot{\gamma} = \omega_x - \tan\theta\left(\omega_y \cos\gamma - \omega_z \sin\gamma\right)$$
$$\dot{\theta} = \omega_y \sin\gamma + \omega_z \cos\gamma \quad (6)$$
$$\dot{\psi} = \omega_y \frac{\cos\gamma}{\cos\theta} - \omega_z \frac{\sin\gamma}{\cos\theta}$$

Where $\gamma$, $\theta$ and $\omega$ are rotation angle, pitch angle and side slip angle.

When $=\pm 90°$, $\tan=\pm\infty$, singular point would appear to $\dot{\gamma}$ and $\dot{\psi}$. In order to resolve this problem, quaternion is adapted. The relationship between quaternion and direction cosine can be showed by the following equation group:

$$\left.\begin{aligned}
l_{11} &= q_0^2 + q_1^2 - q_2^2 - q_3^2 \\
l_{12} &= 2(q_1q_2 + q_0q_3) \\
l_{13} &= 2(q_1q_3 - q_0q_2) \\
l_{21} &= 2(q_1q_2 - q_0q_3) \\
l_{22} &= q_0^2 - q_1^2 + q_2^2 - q_3^2 \\
l_{23} &= 2(q_2q_3 + q_0q_1) \\
l_{31} &= 2(q_0q_2 + q_1q_3) \\
l_{32} &= 2(q_2q_3 - q_0q_1) \\
l_{33} &= q_0^2 - q_1^2 - q_2^2 + q_3^2
\end{aligned}\right\} \quad (7)$$

And the relationship between the quaternion and attitude Euler angle can be expressed by (8)

$$\left.\begin{aligned} q_0 &= \cos(\gamma/2)\cos(\theta/2)\cos(\psi/2) - \sin(\gamma/2)\sin(\theta/2)\sin(\psi/2) \\ q_1 &= -\cos(\gamma/2)\sin(\theta/2)\sin(\psi/2) - \sin(\gamma/2)\cos(\theta/2)\cos(\psi/2) \\ q_2 &= -\cos(\gamma/2)\cos(\theta/2)\sin(\psi/2) - \sin(\gamma/2)\sin(\theta/2)\cos(\psi/2) \\ e_3 &= -\cos(\gamma/2)\sin(\theta/2)\cos(\psi/2) + \sin(\gamma/2)\cos(\theta/2)\sin(\psi/2) \end{aligned}\right\} \quad (8)$$

Following is the calculation method for the quaternion by using aircraft body axis angular velocities $\omega_{xb}$, $\omega_{yb}$ and $\omega_{zb}$.

$$\begin{bmatrix} dq_0/dt \\ dq_1/dt \\ dq_2/dt \\ dq_3/dt \end{bmatrix} = \frac{1}{2} \begin{bmatrix} 0 & -\omega_{xb} & -\omega_{yb} & -\omega_{zb} \\ \omega_{xb} & 0 & \omega_{zb} & -\omega_{yb} \\ \omega_{yb} & -\omega_{zb} & 0 & \omega_{xb} \\ \omega_{zb} & \omega_{yb} & -\omega_{xb} & 0 \end{bmatrix} \begin{bmatrix} q_0 \\ q_1 \\ q_2 \\ q_3 \end{bmatrix}. \quad (9)$$

After obtaining the quaternion, attitude angles are computed by the following equation group.

$$\begin{aligned} \sin\theta &= 2(q_1 q_2 + q_0 q_3) = l_{12} \\ \sin\gamma/\cos\gamma &= -2(q_2 q_3 - q_0 q_1)/\left[1 - 2(q_1^2 + q_3^2)\right] = -l_{32}/l_{22} \\ \sin\psi/\cos\psi &= -2(q_3 q_1 - q_0 q_2)/\left[1 - 2(q_2^2 + q_3^2)\right] = -l_{13}/l_{11} \end{aligned} \quad (10)$$

The aircraft body axis system and the local rectangular coordinate system can be transformed by matrix (11).

$$\begin{bmatrix} x_b \\ y_b \\ z_b \end{bmatrix} = \begin{bmatrix} \cos\theta\cos\psi & \sin\theta & -\cos\theta\sin\psi \\ -\sin\theta\cos\psi\cos\gamma + \sin\psi\sin\gamma & \cos\theta\cos\gamma & \sin\theta\sin\psi\cos\gamma + \cos\psi\sin\gamma \\ \sin\theta\cos\psi\sin\gamma + \sin\psi\cos\gamma & -\cos\theta\sin\gamma & -\sin\theta\sin\psi\sin\gamma + \cos\psi\cos\gamma \end{bmatrix} \begin{bmatrix} x \\ y \\ z \end{bmatrix} \quad (11)$$

The attack angle, side slip angle and rate of change are computed by (12)

$$\begin{bmatrix} V \\ \alpha \\ \beta \end{bmatrix} = \begin{bmatrix} \sqrt{v_{bx}^2 + v_{by}^2 + v_{bz}^2} \\ \arctan\dfrac{-v_{by}}{v_{bx}} \\ \arctan\dfrac{v_{bz}}{\sqrt{v_{bx}^2 + v_{by}^2}} \end{bmatrix} \Rightarrow \begin{bmatrix} \dot{V} \\ \dot{\alpha} \\ \dot{\beta} \end{bmatrix} = \begin{bmatrix} \dfrac{v_{bx}\cdot\dot{v}_{bx} + v_{by}\cdot\dot{v}_{by} + v_{bz}\cdot\dot{v}_{bz}}{\sqrt{v_{bx}^2 + v_{by}^2 + v_{bz}^2}} \\ \dfrac{v_{by}\cdot\dot{v}_{bx} - v_{bx}\cdot\dot{v}_{by}}{v_{bx}^2 + v_{by}^2} \\ \dfrac{\dot{v}_{bz}\cdot(v_{bx}^2 + v_{by}^2) - v_{bz}\cdot(v_{bx}\cdot\dot{v}_{bx} + v_{by}\cdot\dot{v}_{by})}{(v_{bx}^2 + v_{by}^2 + v_{bz}^2)\cdot\sqrt{v_{bx}^2 + v_{by}^2}} \end{bmatrix} \quad (12)$$

In aerodynamics, angle of attack, usually denoted as $\alpha$, specifies the angle between the chord line of the wing of affixed-wing aircraft and the vector representing the relative motion between the aircraft and the atmosphere. Since a wing can have twist, a chord line of the whole wing may not be definable, so an alternate reference line is simply defined. When the component of relative velocity along $y_b$ axis is positive, angle of attack is positive, otherwise contrary.

Angle of side slip relates to the rotation of the aircraft centerline from the relative wind. In flight aerodynamics, this angle is usually denoted as $\beta$ and is usually assigned to be positive when the relative wind is coming from the right of the nose of the airplane. The sideslip angle is essentially the directional angle of attack of the airplane.

## 4    3D Modeling and Visual Scene Driving

### 4.1    3D Modeling

Creator is a real-time 3D modeling tool suite which can generate models with high verisimilitude rate. It integrates polygon mesh modeling, vector modeling and terrain generation into a software package and is mainly applied in visual simulation and interactive simulation. Its powerful modeling functions can provide tools for pattern generator. The advanced real-time functions of this software, such as LOD, polygon screening, logic screening, drawing priority and separation surface, make the OpenFlight (.flt) file the most popular image format in real-time 3D field[6~7].

In this paper, the domestic Kitty Hawk 500 aircraft is used for 3D modeling. Generated model has the same scale parameters as actual Kitty Hawk 500. Dynamic modeling of landing gear, propeller and cockpit door needs the DOF technology provided in Creator in order to display not only flight motion but also motions of some aircraft components. Fig.3 Kitty hawk model and database hierarchy [6][7].

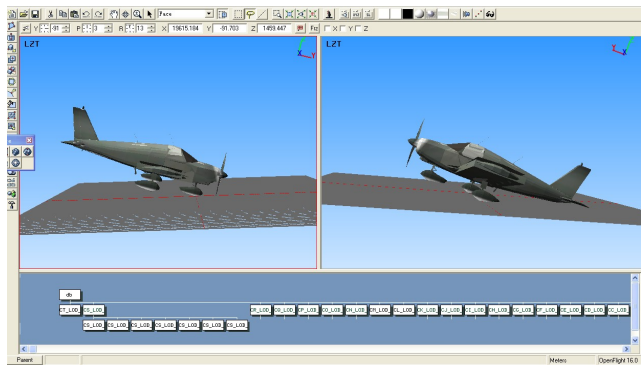

**Fig. 3.** Kitty hawk model and database hierarchy

### 4.2 Model Optimization

In visual simulation systems, loading capacity and rendering efficiency are directly affected by model size, model quantity and range of region terrain. Model optimization is an important step during the design and implementation process because optimized model can provide better authenticity and real-time display. Following optimization techniques are applied for 3D modeling in this paper.

**Removing redundant triangles and polygons.** If too many triangles and polygons existed in one scene, the memory would possibility be exhausted and the acceptable lowest frame rate would not be maintained. This problem should be considered in the beginning of design; and models of realistic facade and simple structure are welcome for FVSS .

**LOD.** Levels of detail (LODs) can be used to stay within the polygon budget and increase viewing performance. LODs are versions of the same model with different numbers of polygons. As the eye point moves closer to the model, more detailed versions are substituted. The version with the highest number of polygons, called the highest LOD, is only displayed when the eye point moves closest to the model in the runtime. When the eye point moves out of the model, not as much detail needs to be visible, a lower LOD is switched in.

**Using texture mapping.** Texture Mapping is the process of applying a texture pattern onto one or more polygons (faces). The Texture Mapping palette is useful for storing a previously mapped texture and its coordinates. You can quickly apply this texture mapping to many faces. This method is useful when you have a large area to cover with a specific texture. Instead of selecting each face and using a Put Texture tool to choose the coordinates for mapping the texture, which can be time-consuming for large databases, you can choose a texture in the Texture Mapping palette with its predefined alignment and select the faces on which to apply it.

**Using instances.** In complex scenes, large amount of memory space would be occupied due to the usage of massive identical geometric solids. Instances can be used to overcome this plight. An instance is a reference to geometry in database file is referenced instead of duplicated. Only one copy of an object's geometry is stored in memory. Instances seem like many shadows of a geometric solid. These instances all have the same properties except space position. The advantages of using instances are that we can save disk space, save time creating new models, and save time editing all occurrences of a model. A possible disadvantage is that you have multiple objects that look exactly the same. This may not be noticeable in large, complicated databases but may be a disadvantage in small databases in which we need more flexibility.

**Creating Billboards.** Billboards are single polygons that are often used to represent symmetrical objects, such as poles, trees, or people. You can apply texture to billboards to create very detailed objects with a low polygon count. Billboards rotate to face the eye point in the runtime system, so the textured side of the polygon is always visible.

**External references.** An external reference is a reference to geometry in another database. With external references, we can reduce the number of polygons in a large database. They are useful for large or complicated databases that use the same geometry in different areas. To create an external reference, we place an external reference node in the database hierarchy, and then assign the directory path and file name of another database to the node. The system uses this directory path to load each external reference. Generally, when loading the program, if all objects were contained in main file, all these objects would be loaded. However, if external reference is used, only objects which fall into sight are loaded rather all objects.[8].

### 4.3    Scene Driving

After the generation of scene models, real-time rendering needs to be performed in Vega Prime which consists of a graphical user interface called LynX Prime and Vega Prime libraries and header files of C++-callable functions. LynX Prime is the graphical user interface for defining, editing, and previewing Vega Prime applications. During the runtime of the application, selection function in user interaction is completed by Vega Prime library function calls.

A convenient class called vpApp is shipped with Vega Prime. vpApp is constructed based on the initialization, define, configuration, frame loop, and shutdown flow. The vpApp is used in all code examples through out this guide [9]. A basic Vega Prime application will look like this:

```
#include <vpApp.h>
int main(int argc, char *argv[])
{
  // initialize vega prime
  vp::initialize(argc, argv);
  // create a vpApp instance
  vpApp *app = new vpApp;
  // load acf file
  if (argc <= 1)
  app->define("kitty hawk500.acf");
  else
  app->define(argv[1]);
  // conFig. my app
  app->conFig.();
  // frame loop
  app->run();
 // unref my app instance
  app->unref();
```

```
    // shutdown vega prime
    vp::shutdown();
    return 0;
}
```

### 4.4   Terrain Modeling

Terrain modeling is the generation of a polygon set with which these polygons has certain organized sequences and can approximately represents surface conditions of earth. During the generation, DEM of realistic terrain with appropriate proportion, geomorphology feature data and appropriate transition algorithm are used [10]. In this paper, the terrain of Shenyang region is based for terrain modeling. And generated models are used for terrain models in system scenes. Terrain modeling contains the following main steps:

**Get terrain DEM data.** The DEM data file (.img format) for Shenyang region (north latitude 41.8°, east longitude 123.4°) is downloaded from International Science Data Services Platform;

**Data conversion.** The downloaded .img file is converted to .DEM file by Global Mapper software, and the .DEM file is continuously converted to .DED format through the command "readusgs" in Creator;

**Create terrain file.** Open the Terrain tool in Creator and import the .DED file created in step b, split the DED file into partitions by applying the technology of Large Area Database Management. Through Page scheduling for these partitions, memory usage and rendering time would be decreased. By partitioning DED file, large terrain can be conveniently integrated into the program.

## 5   Experiment Analyses

The simulation result is verified in Matlab. Parameters of Kitty Hawk 500 are used for the simulation of flight motion equations. These parameters include mass (1000kg), flight velocity (250km/h), flying altitude (3000m), mean aerodynamic chord (1.9m), wingspan (10m), yaw angle (0deg/s), pitch angle (0deg/s) and roll angle (0deg/s). Fig.4 shows the changing curve of relative air speed, Mach number and dynamic pressure in short period; Fig.5 shows the changing curve of axial velocity; and the changing curve of yaw angle, pitch angle and roll angle is shown in Fig.6. As we can see, the Matlab simulation results are consistent with theory analysis and show higher simulation accuracy.

Under above conditions, the FVSS is simulated and the results are shown in Fig.7 and Fig.8. Fig.7 is a flying screenshot of Kitty Hawk 500 which is 3000 meters above Shenyang region. Fig.8 is a scene looked from cockpit. Seen from the experiment results, it can be concluded that the FVSS developed through integrated development platform are capable of obtaining better fidelity and real-time.

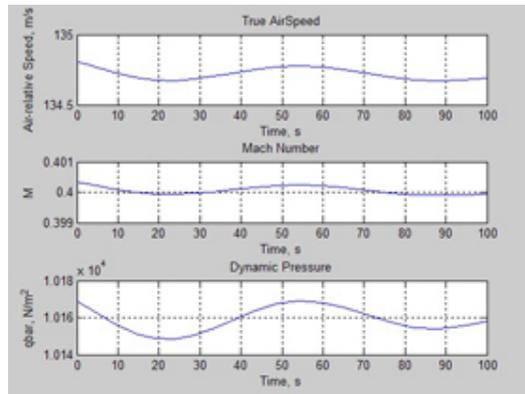

**Fig. 4.** The changing curve of relative air speed, Mach number and dynamic pressure in short period

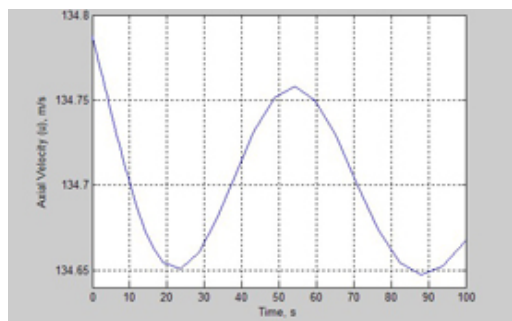

**Fig. 5.** The changing curve of axial velocity

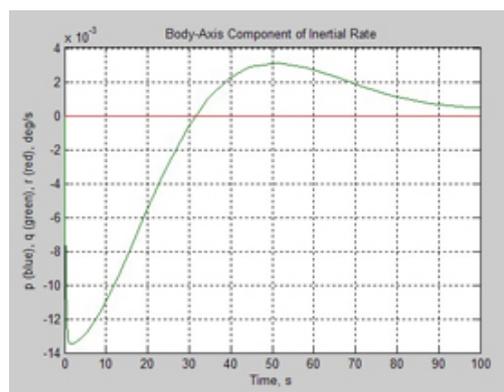

**Fig. 6.** The changing curve of yaw angle, pitch angle and roll angle

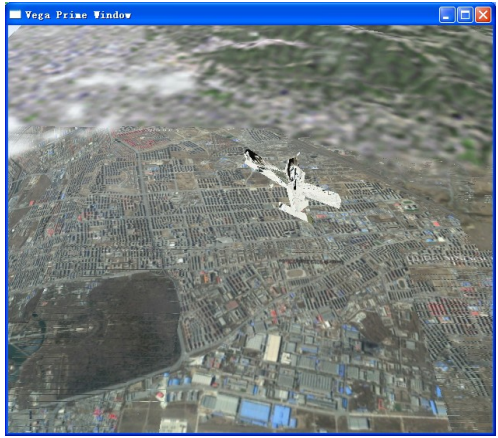

**Fig. 7.** A flying screenshot of Kitty Hawk 500 which is 3000 meters above Shenyang region

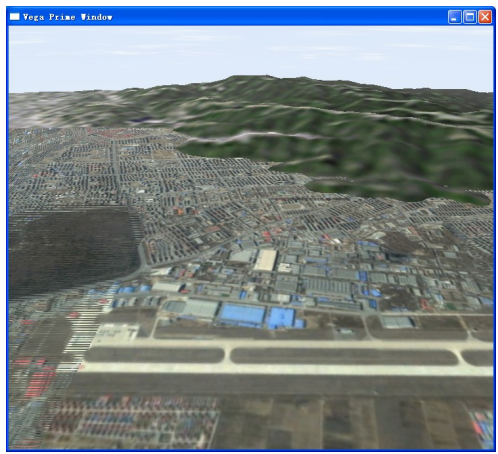

**Fig. 8.** A scene looked from cockpit

### References

1. Sun liqing, Zhang guofeng, Wang xingren: Unmanned Areial Vehicle Simulation Training System. Computer Simulation. 23, 44--46 (2006)
2. Peng liang, Huang hanxin: Simulation system of cruise missile using VC and Vega Prime. Journal of Central South University (Science and Technology). 41, 219--222 (2010)
3. Lu ping. Research on Visual Simulation System of Flight Simulator. Jilin Universtiy. (2008)
4. Lu ming. Design of Primary Training Aircraft Flight simulator. Liaoning Technical Universtiy. (2009)

5. Fan jun. The Design and Simulation of Training Flight simulator system. Northwestern Polytechnical Universtiy. (2007)
6. Multi Gen-Paradigm Inc. Creator Terrain Studio Desktop Tutor. (2006)
7. Multi Gen-Paradigm Inc. Creator Terrain Studio User's Guide. (2006)
8. Wang Mingyin, Wei qun, Xu en, He guolin: Research on Key Techniques of Virtual Reality for Large Scenes Based on Creator/Vega Prime. Journal of System Simulation. 21, 117--120 (2009)
9. Multi Gen-Paradigm Inc. Vega Prime Programmer's Guide. (2006)
10. Multi Gen-Paradigm Inc. Vega Prime Options Guide. (2006).
11. Wu Xiaojun, Wang Changjin: Realtime Simulation of Battlefield Fight Scene System Based on Creator/Vega. Journal of System Simulation. 41, 219--222 (2010)
12. Cai zhongfa, Liu dajian, Zhang anyuan: Scheme Study of Automobile Driving Training Simulator Based on Virtual Reality. Journal of System Simulation. 14, 771--774 (2002)
13. Song zhiming, Kang fengju, Nie weidong et al: Study on the Method for Introducing Input Device in Vega Application. Ship Electronic Engineering. 24, 96--98 (2004)
14. Zhang zhili, Chen shan, Long yong, Wang haiyan: Development of Missile Simulation Training System Based on HLA. Comupter Simulation. 24, 75--84 (2007)
15. Zhang lin, Yang zhaosheng, Wang bin, Hu juanjuan, C.: Study on 3D-map for On-board Navigation System Based on Multigen Creator & Vega. In: Intelligent Transportation Systems Conference (ITSC 2006), pp. 933--938. IEEE Press, New York (2006)